\documentclass[12pt]{iopart}

\usepackage{latexsym}
\usepackage{amsfonts}
\usepackage{epsfig}
\usepackage{amssymb}
\usepackage{mathrsfs}
\usepackage{graphicx}% Include figure files
\usepackage{dcolumn}% Align table columns on decimal point
\usepackage{bm}% bold math
\begin{document}

\title{Water at interface with proteins}

\author{Giancarlo Franzese$^1$, Valentino Bianco$^1$ and Svilen Iskrov$^{1,2}$}

\address{$^1$Departament de F\'{\i}sica Fonamental, Universitat de Barcelona\\
Diagonal 647, 08028 Barcelona, Spain\\
$^2$Ecole Normale Sup\'erieure de Cachan, Paris}

\ead{gfranzese@ub.edu}

\bigskip

\begin{abstract}
Water is essential for the activity of proteins. However, the effect
of the properties of water on the behavior of proteins is only
partially understood. Recently, several experiments have investigated
the relation between the dynamics of the hydration water and the
dynamics of protein. These works have generated a large amount of data
whose interpretation is debated. New experiments measure the dynamics
of water at low temperature on the surface of proteins, finding a
qualitative change (crossover) that might be related to the slowing
down and stop of the protein's activity (protein glass transition),
possibly relevant for the safe preservation of organic material at low
temperature. To better understand the experimental data several
scenarios have been discussed. Here, we review these experiments and
discuss their interpretations in relation with the anomalous
properties of water. We summarize the results for the thermodynamics
and dynamics of supercooled water at an interface. We consider also
the effect of water on protein stability, making a step in the
direction of understanding, by means of Monte Carlo simulations and
theoretical calculations, how the interplay of water cooperativity and
hydrogen bonds interfacial strengthening affects the protein cold
denaturation.
\end{abstract}

\vspace{2pc}
\maketitle

\section{Introduction}

Water is ubiquitous is biological systems. It is a major component of
cells and participates in the majority of the biological
processes. It is usually considered essential for life, but it is still
under debate why \cite{life}. One possible reason is that water has
many properties that are unusual with respect to other liquids \cite{FS2007}. 

The anomalous behavior of water is evident in the liquid phase.
For example, fluctuations of volume and fluctuations of entropy have a
minimum for liquid water, while in usual liquids they decrease when
the temperature $T$ is decreased. 
Volume fluctuations can be observed by measuring the 
compressibility $K_T$, defined as how much the
volume changes when the pressure $P$ is changed at constant $T$, 
and entropy fluctuations are proportional to the specif heat $C_P$ at
constant $P$. For water at ambient pressure 
$K_T$ has a minimum at $46^\circ$C and $C_P$ at 
minimum  at $35^\circ$C.

The anomalies of water become more evident when $T$ 
is decreased toward and below $0^\circ$C.
For example, water has a maximum in density at $4^\circ$C.
Normal
liquids, such as argon, reduce their density when the temperature
decreases and reach their maximum density when they solidify in a
crystal. Water, instead, below $4^\circ$C
expands. Therefore, liquid water at $0^\circ$C has
a density smaller than water at $4^\circ$C. By solidifying into ice, water
expands even further, becoming ``lighter''. For this reason ice cubes
float in a glass of water. This property  has a dramatic
consequence in processes such as the cryopreservation of biological
cells, because the large amount of water in each cell expands when
forms ice and breaks the cell. 

Another property of water that is relevant in these conditions is that
water can stay in its liquid state even at $T<0^\circ$C, i.e. it 
can be supercooled below its melting temperature. Bulk water 
can be supercooled to  $-41^\circ$C at atmospheric pressure, but
in different conditions water can remain liquid even at lower
temperatures. For example, it can be supercooled  down to
$-47^\circ$C when confined in vegetable fibers, or down to $-92^\circ$C 
when compressed at $2$~kbars. An anti-freezing effect can be
achieved also by dissolving in water polymers or proteins, with several
practical applications.  Understanding the details of
this phenomena and how to regulate it could be very relevant for 
cryopreservation, food storage and refrigeration \cite{Sun2003,Petzold2009}.

Also in its solid state, water is peculiar. Water is a polymorph
with many different crystal phases, more than fifteen, some of which
are stable only at pressure $P>100$~GPa. But water can easily form a
solid that is not a crystal. If quenched rapidly below $-123^\circ$C
at ambient pressure, liquid water freezes in a metastable
amorphous state, which is an arrested liquid configuration \cite{FS2007}. 
At low pressure water forms a low-density
amorphous (LDA) state \cite{BrugellerMayer1980}, while at high
pressure it forms a high-density amorphous (HDA) state
\cite{Mishima1985}, separated by a volume
discontinuity of $\approx 27$\%, comparable to that between
crystalline ice I and ice VI. A smaller discontinuity has been
observed more recently \cite{Loerting01,Finney02}, but its
interpretation is under debate.

The discontinuities between amorphous states 
at different densities and the fact that quantities
such as $K_T$ or $C_P$ largely increase in the supercooled state, show
that water has a complex behavior at low $T$. This observation and the
fact that water remains in its liquid state at very 
low $T$ when in contact with 
organic or inorganic interfaces suggest that
water could play a main role in phenomena such as the so-called
protein glass transition, or the protein cold denaturation at low temperature.

\subsection{Experiments}

To explore how the dynamics of  proteins and water are related,
S-H. Chen et al. in 2006 studied by high--resolution
quasi--elastic neutron scattering (QUENS) the structure and dynamics of water
molecules in the hydration layer surrounding lysozyme proteins at
temperatures around $220$~K ($-53.15^\circ$C) \cite{Chen2006}. Below
this temperature the protein is in a solid--like ``glassy state'', with no
conformational flexibility and  no biological functions. As the
temperature is increased, the protein displays a harmonic 
atomic motion that, in hydrated proteins, suddenly becomes anharmonic
and liquid-like at about $220$~K. 
The change in the protein dynamics is believed to be triggered by the
coupling with the hydration water through the hydrogen bonds 
because the hydration water displays a dynamic transition at a similar
$T$ \cite{Paciaroni1999,Caliskan2002}. Chen interpreted the hydration
water dynamic transition as a change in a main structural relaxation
and recently extended this interpretation to hydrated proteins at
high-pressure.

This explanation has been questioned by Swenson et
al. \cite{Swenson2008}. 
By using dielectric measurements on
myoglobin in water-glycerol mixtures, they found a dynamic crossover
at about $200$~K and they interpreted it
as an evidence that local (secondary)
protein motions are controlled (slaved) by the local fluctuations in the hydration shell,
as proposed by Fenimore et  al. in 2004 \cite{Fenimore2004} based on
M\"ossbauer and neutron-scattering experiments. 

On the other hand, Pawlus et al. in 2008, based on measurement of  conductivity on
hydrated lysozyme, found no
crossover around $220$~K. They also ascribe the apparent crossover
observed with QENS  to a secondary relaxation and  
to a lack of resolution on main structural relaxation of
QENS \cite{Pawlus2008}. 

More recently nuclear magnetic resonance (NMR) experiments on water
hydrating elastin and collagen, 
performed by  Vogel, showed no crossover at $220$~K but 
a crossover at about $200$~K.
This data have been interpreted as consistent with thermally
activated tetrahedral jump motion of hydration water \cite{Vogel2008}.

These and other experiments, therefore, show that it is difficult to
achieve a clear understanding of the protein-hydration water coupling
solely based on the experimental data.
A possible way to gain further insight is offered by
the numerical results of detailed computer simulations. 

\subsection{Numerical results}

In 2006 Kumar et al. \cite{Kumar2006}, by simulating lysozyme or DNA 
surrounded by water represented with the TIP5P model, found a dynamic
transition of the macromolecules at about the same temperature of a
dynamic crossover in the diffusivity of hydration water, in the range
$242$~K--$250$~K. They also show numerical indications that  this
crossover coincides with the maximum of the isobaric specific heat of
the whole system and the maximum fluctuations in tetrahedral order of
the hydration water. 

This observation was confirmed in 2008 by Lagi et
al. \cite{Lagi2008}. They found a 
strong crossover in the water translational ($\alpha$) relaxation time and
in the inverse of its self-diffusion constant  
at about $223$~K, by simulating hydrated lysozyme with
water represented by the TIP4P-Ew model. They corroborate that the
crossover  corresponds to the maximum structural change, and they
found a low activation energy, showing consistency with the neutron
scattering and the NMR experiments.

Nevertheless, in 2009 this view was disputed by Vogel
\cite{Vogel2009}. By simulating elastin or collagen hydrated by
SPC-water he found only a weak crossover in the hydration water
correlation time at about $200$~K and a  
 high activation energy associated to a secondary relaxation,
 consistent with activation energies determined in dielectric
 spectroscopy  and NMR studies. 

It is evident from these and other  numerical studies that
the comparison of experiments with simulation is useful, but is not
enough to elucidate the mechanisms that regulate the coupling of the
water dynamics with the biomolecules dynamics. In particular, the
difficulties in finding clear answers to the open questions rise for
the fact that experiments and numerical results are both affected
by errors and uncertainties. Therefore, it is natural to look for a
theory that could be able to find exact relations and make predictions
to test in further experiments. 

In the next section we describe a model for a hydration water
monolayer that allows to develop a theory for these phenomena and to
predict properties that could be verified in experiments. The model,
moreover, allows for efficient simulations whose results complement
the theoretical analysis.  In section 3 we review results of this
tractable model, including new data for the density and energy
distributions at different pressures and very low temperature
(subsection 3.3) and for the hydration percolation (subsection
3.4). In section 4 we discuss new results about protein stability
and confined water in the context of food processing and cells in
living organism.  In section 5 we give our conclusive remarks and we
discuss possible extensions of this model.

\section{A tractable model for a hydration water monolayer}

We first consider the case of water nanoconfined between two
hydrophobic surfaces. To fix the idea, let's consider the case with a 
distance $\delta=0.7$~nm between the two surfaces. Because
the confining surfaces are hydrophobic, the water molecules will not
form hydrogen bonds (HB) with the surfaces. Experiments shows
that in bulk each water molecule is surrounded by four nearest
neighbor molecules at a distance of about $3$\AA~ with a structure that
resembles that of a tetrahedron at low $T$V and $P$
\cite{Soper-Ricci-2000}.  Hence, one would expect that each water molecule
between the two hydrophobic plates will adjust in a way to form a distorted
network of HBs with each molecules surrounded by other four. This has been
indeed found by Kumar et al. \cite{Kumar2005} by simulating
TIP5P-water in these conditions.

\begin{figure}
\begin{center}
\includegraphics[width=4cm]{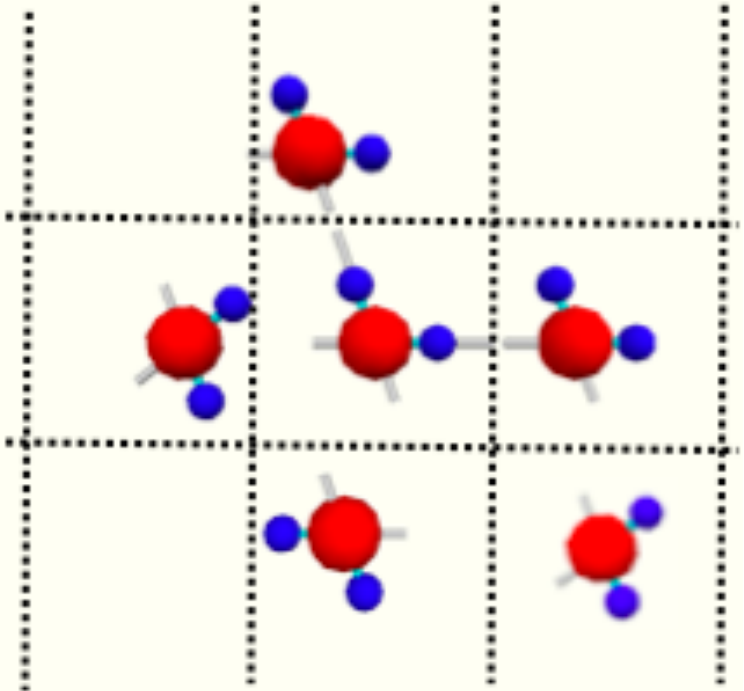}
\end{center}
\caption{Schematic representation of a water monolayer. Top view of water
  molecules with an oxygen atom (red) and two hydrogen atoms (blue)
  distributed  over a  surface, between two hydrophobic plates
  (not represented). Possible hydrogen bonds are represented by gray
  sticks.  The total surface area is divided in equal-size
  square cells (dashed lines). In the tractable model adopted here,
  the coordinates of each molecule inside a cell are
  coarse-grained. A configuration of water molecules is
  represented by the occupancy state of the cell (local density) and
  the states of the four bonding indices of each molecule, accounting
  for the hydrogen bonds formed with water molecules in the nearest
  neighbor cells. }
\label{schematic}
\end{figure}

To define a tractable model
\cite{FYS2000,FS-PhysA2002,FS2002,FMS2003,FS2005,FS2007}, we
coarse-grain the structure described 
above, dividing the slab of space occupied by water
into cells with volume $v=\delta r^2$, with a square section of size $r\geq
r_0$, where $r_0=2.9$\AA~\cite{Narten67} is the closest approach (van
der Waals) distance between water molecules (Fig.~\ref{schematic}). If the density of water
is $\rho$, and the number of water molecules in $N$, the water volume
is $V\equiv N/\rho$. If the system is uniform and we divide the system in $N$
cells, each cell has on average one water molecule. 
More in general if the system is not uniform, some cells can be empty or
each cell can occupy a different
volume $v_i$ and the distance $r_{ij}$ between two molecules in the cells $j$
and $j$ is the distance between their centers. For example, if $i$ and
$j$ are the indices of nearest neighbor cells, the distance between
the molecules in these cells
is $r_{ij}\equiv (\sqrt{v_i/\delta}+\sqrt{v_j/\delta})/2$ .
 
We describe the isotropic (e.g., van
der Waals) attractive and repulsive interactions between the molecules
by  a standard a Lennard-Jones interaction 
\begin{equation}
U\equiv \sum_{ij}
\epsilon\left[\left(\frac{r_0}{r_{ij}}\right)^{12}-
      \left(\frac{r_0}{r_{ij}}\right)^6\right] 
\label{U0}
\end{equation}
where $\epsilon=5.8$~kJ/mol~\cite{Henry2002} is the attractive energy
and the sum is 
over all the possible pairs of molecules $i$ and $j$.
Some modifications of this
interaction, such as the introduction of a maximum cut-off distance or
a hard-core distance, have been adopted in previous analysis of this
model
\cite{FMS2003,KFS2008,FDLS2009,dlSantos09,MSSSF2009,Franzese2010,Stokely2010} 
to simplify the numerical simulations. For the theoretical analysis,
instead, this interaction term has been replaced by a more tractable
discrete interaction \cite{FYS2000,FS-PhysA2002,FS2002,FMS2003,FS2005,FS2007,FranzeseSCKMCS2008,KumarFS2008,Stokely2010}.

To take into account the energy and entropy variation when water
molecules form HBs, we introduce for each water molecule $i$ four
bonding indices $\sigma_{ij}$, one for each of the possible HBs
with the four nearest neighbor water molecules $j$. The index can assume
$q$ different states, $\sigma_{ij}=1, \dots , q$, where $q=6$ is a
parameter whose value we will discuss in the following. Therefore,
each water molecule has $q^4=6^4=1296$ possible states, that can be
interpreted as possible rotational configurations. The total
number of configurations for the system is $q^{4N}$, that for $N=10^5$ (the
maximum number that we have considered in our analysis) is an
astronomical number (about $3\times 10^{311260}$).
When a molecule forms a
HB the number of its accessible configurations decreases, and the 
energy of the system is reduced. 

To estimate the decrease in the number of
accessible configurations, we observe that a HB is broken if it
deviates from a linear bond more than $\pm 30^\circ$. Therefore, only 
$1/6$ of the whole
continuous range of orientations $[0,360^\circ]$ in the OH---O plane 
are
associated to a bonded state. Hence, $5/6$ of the possible
configurations are not--bonded. By allowing $q=6$ possible states for each 
bonding index $\sigma_{ij}$, we can count correctly the entropy loss
associated to the formation of a HB if only one out of 
$q$ states corresponds to a HB. This is achieved by allowing the formation of the HB between
molecules $i$ and $j$ only if $\delta_{\sigma_{ij},\sigma_{ji}}=1$, 
where by definition $\delta_{a,b}=1$ if $a=b$ and $\delta_{a,b}=0$
otherwise.

Next, we take into account that 
a HB is broken if the OH---O
distance is too large
\cite{Khan2000,MartinsCosta2005}. To simplify the analysis, we
introduce a condition on the O--O distance $r$, allowing the formation
of HB only if $r\leq r_{\rm max}$, with 
$r_{\rm max}=r_0\sqrt{2}=4.10$\AA. Therefore, considering that the
length of the OH covalent bond is $r_{\rm
  OH}=0.96$\AA~\cite{Csaszar2005}, 
the maximum bond length is
$r_{\rm max}-r_{\rm  OH}=3.14$\AA, consistent with other
choices in literature \cite{Khan2000,MartinsCosta2005}.
In particular, we impose this condition by introducing a discrete
variable $n_i$ for each cell $i$, with $n_i=1$ if $r_i\leq r_0\sqrt{2}$,
otherwise $n_i=0$. A cell with $n_i=1$ is liquid-like because its
(dimensionless) density $\rho_i\equiv (\delta r_0^2)/(\delta r_i^2)\geq 1/2$, while a cell
with $n_i=0$ is gas-like because has $\rho_i<1/2$. Hence, we define
the number of HBs as 
\begin{equation}
N_{\rm HB}\equiv \sum_{\langle i,j\rangle} n_in_j \delta_{\sigma_{ij},\sigma_{ji}} 
\label{NHB}
\end{equation}
where the symbol 
$\sum_{\langle i,j\rangle}$ denotes that the sum is performed over 
nearest neighbor cells $i$ and $j$.

From the experiments we know the formation of HBs leads to an open
network of molecules with, on average, four neighbors instead of
twelve as in argon-like fluids. The resulting 
volume per molecule with HBs is larger than the volume
per molecule with no HBs. This is observable as the
anomalous density decrease described in the introduction, that is
a consequence of the formation of a macroscopic number of HBs. 
This effect is incorporated in the model by considering the total
water volume to be given by
\begin{equation}
\label{vol}
V \equiv V_0 + N_{\rm HB} v_{\rm HB},
\end{equation}
\noindent
where $V_0$ is the water volume in absence of HBs, and $v_{\rm HB}$ is
the increase of volume per HB. 
To estimate the parameter $v_{\rm HB}$, we consider as reference
values the increase between the density $\rho_{\rm Ih}=0.92$~g/cm$^3$ 
of the ice Ih at atmospheric pressure and ice VI, with density 
$\rho_{\rm VI}=1.31$~g/cm$^3$, or ice ice VIII, with density 
$\rho_{\rm VIII}=1.46$~g/cm$^3$. Ice Ih is
characterized by
hexagonal rings of HBs with an almost perfect  tetrahedral structure,
while ice VI and ice VIII have 
a structure consisting of two interpenetrating
tetrahedral networks of HBs.
The  relative increase of density in
these cases is
equal to $0.42$ and $0.59$, respectively. Hence, in the model we set the relative
HB increase of volume per molecule equal to the average between the
previous reference values, i.e. $v_{\rm HB}/\delta r_0^2=0.5$.

Note that this increase of volume per molecule is due to a decrease of
first neighbors and does not imply an increase of distance between molecules.
Our model, being coarse-grained, does not include all the details
about the structure, but maintains the increase of volume per molecule
with no effect on the distance $r$ between  molecules. In particular, the HB volume
increase does not affect the calculation of $U_0(r)$ of Eq.(\ref{U0}).

The formation of a HB leads to an energy gain, represented in the
model by an interaction term
\begin{equation}
\label{hb}
\mathscr{H}_{\rm HB} \equiv - J N_{\rm HB} ,
\end{equation}
where $J$
is the characteristic energy of the covalent (directional) component of the HB.
This term only accounts for the two--body component of the HB
interaction. However, in water many--body effects are relevant and,
in particular, the three--body term~\cite{Pedulla1996,Skinner2008}. 
This can be observed from the $T$-dependence of the O--O--O angle
distribution. This distribution becomes sharper around the tetrahedral 
angle when $T$ decreases \cite{Ricci09}.
Hence, we include in the model the many--body (cooperative) effect due to HBs
\cite{Ohno2005,Cruzan1996,Schmidt2007}, which minimizes the 
energy when the HBs of nearby molecules optimize the tetrahedral orientation.
This is accomplished by further adding to the Hamiltonian in
Eqs.~(\ref{U0}) and (\ref{hb}) the term
\begin{equation}
\mathscr{H}_{\rm coop} = -J_\sigma \sum_i n_i \sum_{(k,\ell)_i} \delta_{\sigma_{
ik},\sigma_{i\ell}} ,
\label{coop}
\end{equation}
%\noindent
where $J_\sigma$ is the characteristic energy of the cooperative component of 
the H bond, and the sum over $(k,\ell)_i$ is performed over all
the six different pairs of the four bonding indices of molecule $i$. 

As discussed by Stokely et al.  in Ref.~\cite{Stokely2010},
Eqs.~(\ref{hb}) and (\ref{coop}) have two free
parameters: $J$ and $J_\sigma$, respectively. 
Experiments estimate the HB in ice Ih to be $\approx 3.0$ kJ/mol 
stronger than in liquid water~\cite{newref}. Attributing this increase to a
 cooperative interaction among HBs \cite{Heggie1996}, we can
 estimate the value of $J_\sigma$ in the cell model to be $\approx
 1.0$ kJ /mol, because for each HB there would be $6/2$ pairs of
 $J_\sigma$--interactions. 
The optimal HB energy, $E_{\rm HB}$, has been measured to be
$\approx 23.3$ kJ/mol~\cite{Suresh00}.  By considering tetrahedral
clusters of H-bonded molecules, with HB and van der Waals
interactions up to the third  nearest neighbor molecules, 
the value for the directional component of the HB is estimated by
Stokely et al. \cite{Stokely2010} as $J\approx 12.0$ kJ/mol.
Other experimental estimates suggest that breaking the directional
component of the HB requires $J\approx 6.3$ kJ/mol~\cite{Chumaevskii2003}.
It is, therefore, a reasonable estimate to set $J_\sigma/J=1/10$ and
to consider $J$ as the only free parameter of the model.

In the following we will briefly summarize some recent results for this model,
to show that it reproduces in a qualitative way the properties of
water. An appropriate choice of the free parameter $J$ leads to
results that, as for more detailed models, can be fairly rescaled on
the known proprieties of a water monolayer. In this respect, this model is not
better then detailed models, but, it has two features that
detailed models have not. i) It is less computationally expensive,
because it is coarse-grained. This allows to simulate very large
number of water molecules (about a million) on simple desktop
computers in a few hours. ii) More importantly and differently from
the detailed models, this model is tractable
for theoretical calculations. The trade-off for these advantages is
that the model is coarse-grained and cannot give informations about
some properties of water, such as the structure. Nevertheless, works
is in progress to overcome this limitation. 

\section{Results for water between hydrophobic plates}

We study the model described in the previous section
by mean--field (MF) analysis and Monte
Carlo (MC) simulations. The MF approach follows the Bethe-Peierls and the
cavity method \cite{MezardParisi01}, by expressing
the molar Gibbs free energy in terms of an exact partition function for
a portion of the system, and taking  into account the effect of all the rest of the system as a
mean field acting on the border of this portion, as described in
Ref.~\cite{FS2007,Odessa2009}. 

MC simulations are performed at constant $N$, $P$, $T$, allowing the
volume $V_0$ in Eq.(\ref{vol})  to fluctuate as a stochastic
variable. To minimize the boundary effects, we consider 
periodic boundary conditions in the directions parallel to the
confining surfaces.
To study the thermodynamic properties of the model we adopt
an efficient cluster MC dynamics, defined in
Ref.~\cite{MSSSF2009}, or a continuous $T$ algorithm, the histogram
reweighting method, as in Ref.~\cite{FMS2003}.
To study the dynamics of the HBs we adopt a standard Metropolis
 algorithm \cite{KumarFS2008}, while to study the diffusion properties
 we use the Kawasaki algorithm \cite{FDLS2009}.

\subsection{The gas and liquid phases for nanoconfined water:
  transport properties}

\begin{figure}
\includegraphics[width=8cm]{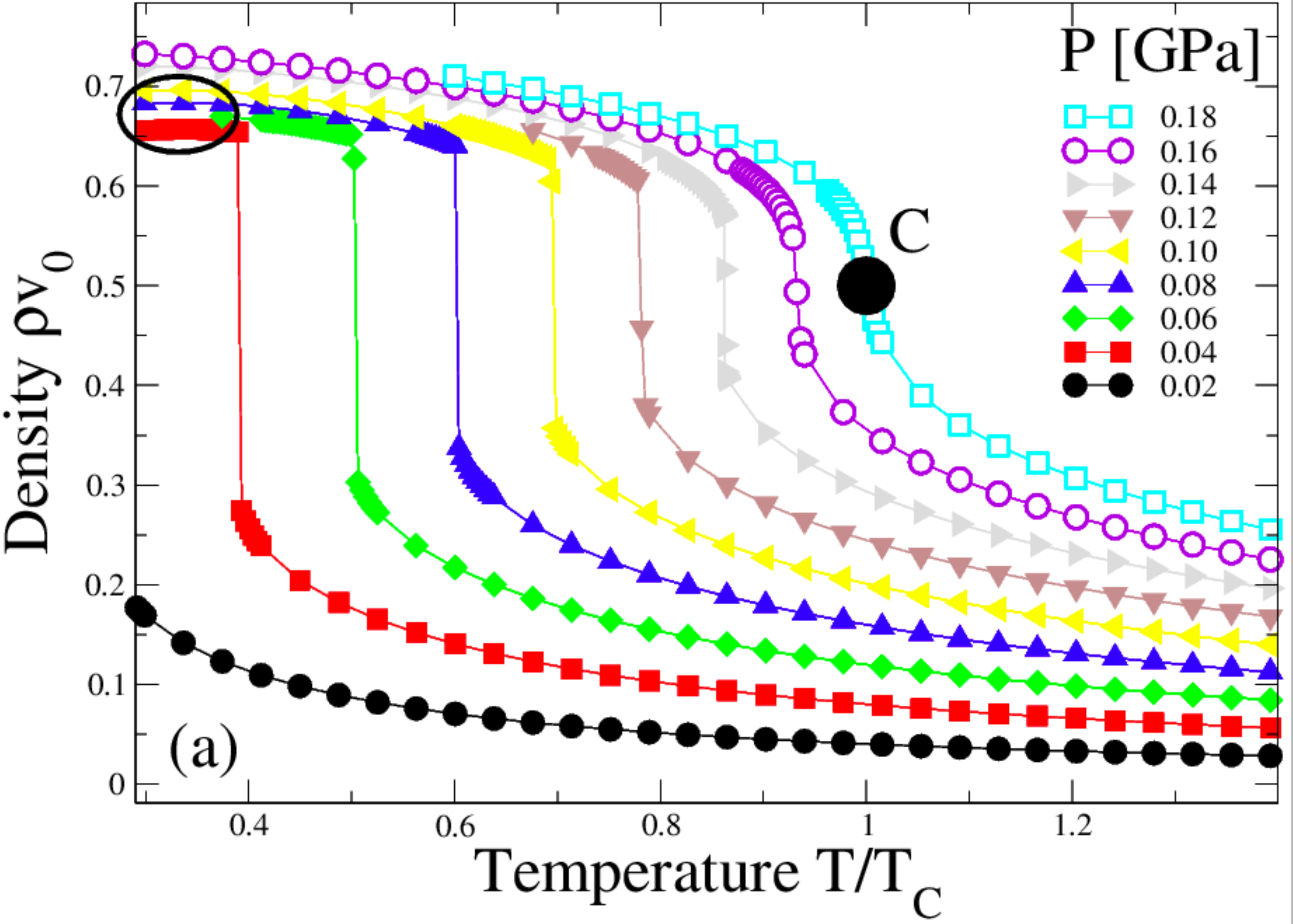}
\includegraphics[width=8cm]{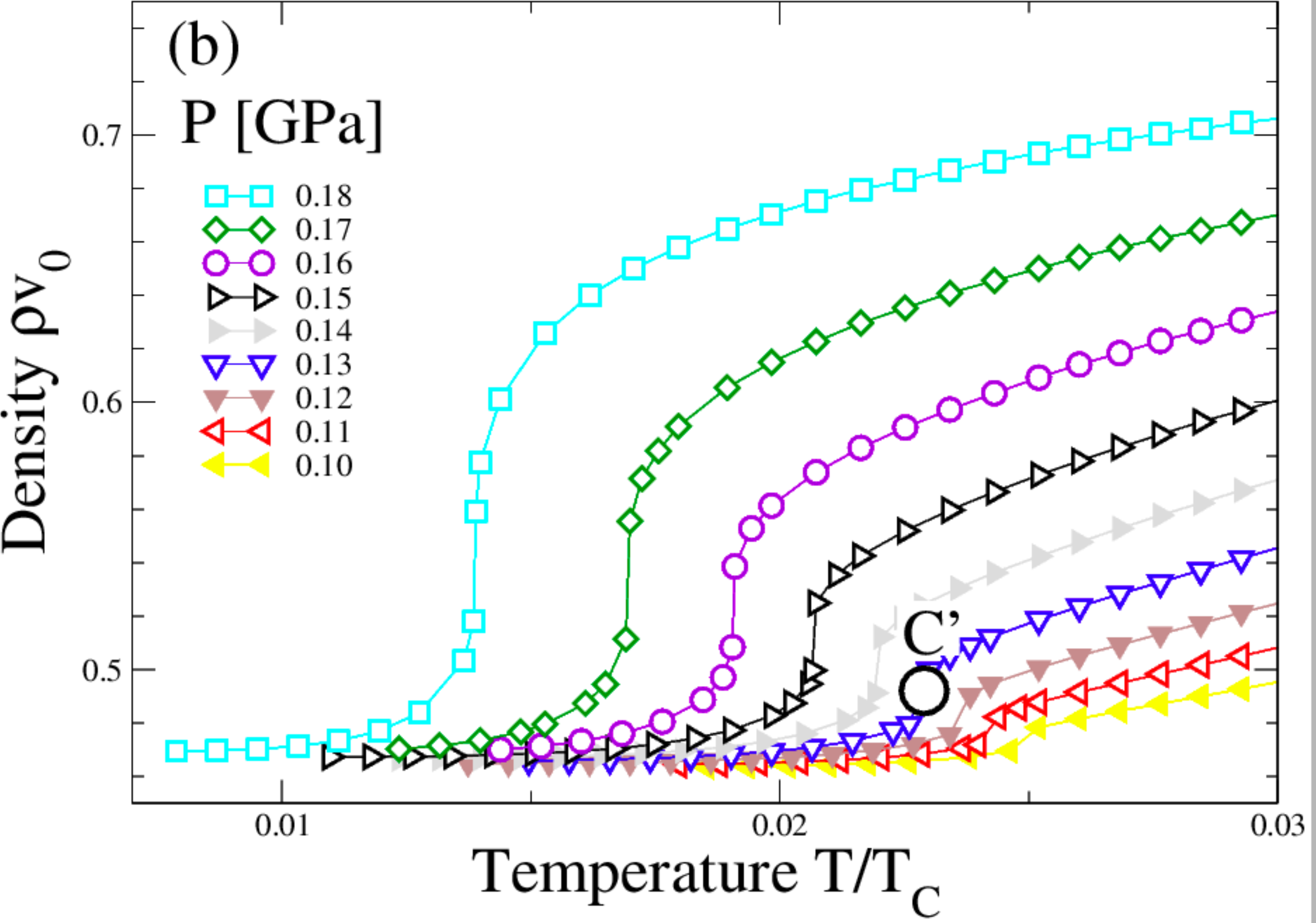}
\caption{Density at constant pressure for a water monolayer
  nanoconfined between hydrophobic slabs of infinite section and
  separated by a distance $\delta=0.7$~nm. Results are from Monte Carlo
  simulations of a system with $N=15625$ water molecules. 
 The pressure $P$ is calculated in GPa, the temperature $T$ is
 rescaled with respect to the liquid-gas critical point temperature
 $T_C$ of the water model, the density $\rho=N/V$ is rescaled by the
volume $v_0=r_0\delta\simeq 0.059$~nm$^3$,
 where $r_0=2.9$\AA~ is the   van der Waals distance.
(a) Isobars at high $T$, around the liquid-gas critical point $C$,
marked by a full large circle. At $P<P_C\simeq 0.18$~GPa a
discontinuity in the isobars denotes the coexistence of the gas (at
$\rho v_0<0.5$) with the liquid (at $\rho v_0>0.5$).
In the circled region for $0.02$~GPa$\leq P\leq 0.06$~GPa
the isobars reach the temperature of maximum density (TMD) at $\rho
v_0\simeq 0.65$ and $T/T_C\simeq 0.35$. 
(b) At much lower $T$, the isobars display another discontinuity in
density that is evident at $P=0.18$~GPa, and smoothly disappears,
within the calculation error,  for $P<0.13$~GPa. This discontinuity
denotes the coexistence of a liquid at higher density with a liquid at
lower density. Where the discontinuity disappears, the system displays
a liquid-liquid critical point $C'$ (large open circle, with
$P_{C'}\simeq 0.13$~GPa and $T_{C'}\simeq 180$~K.
In both panels errors are of the order of 
the size of symbols.}
\label{rho}
\end{figure}

For water nanoconfined between hydrophobic plates, the model display
a gas--liquid first-order phase transition ending in a critical
point $C$, that qualitatively resembles the gas--liquid
transition for bulk water (Fig.~\ref{rho}a). In their work, de los Santos et
al. \cite{dlSantos09,DLSF2010} verify that the water diffusion
constant $D$ decreases when water changes from gas to liquid.
The change in $D$ is strong far from the  critical
point $C$, and it disappears at $C$. In the liquid phase, de los
Santos et  al. \cite{dlSantos09} observe that, as for bulk water, the
nanoconfined system has a region in the $P$--$T$ plane where  $D$
 increases for increasing $P$. This behavior is an anomaly of
water, because in normal fluids $D$ decreases for increasing $P$
\cite{vilaseca1,vilaseca2}.  
This anomalous behavior is qualitatively rationalized as a
consequence of the HB formation and it 
has been observed both in bulk and confined water \cite{Han08}. 

At lower temperature, by decreasing $T$
at constant $P$ the model displays the line of temperatures of maximum
density (TMD) \cite{KumarFS2008,FDLS2009}, as in the bulk case (Fig.~\ref{rho}a).
The TMD line in the $P$--$T$ plane has a positive slope at low $P$ 
and negative slope at high $P$.

The nanoconfinement does not allows the formation of crystal
ice. Nevertheless, it is possible to calculate where in the  $P$--$T$
plane $D(P,T)$ is constant and to show that the
lines of constant-$D$ qualitatively resemble the melting line of bulk
water \cite{dlSantos09,DLSF2010}.

\subsection{Water monolayer compared to protein hydration water}

At low $T$  the diffusion constant largely decreases and,
eventually, the monolayer becomes subdiffusive, i.e. the mean square
displacement of water molecules never reach the diffusive regime
\cite{dlSantos09,DLSF2010}. This property has been observed 
experimentally below $320$~K by neutron scattering in a monolayer of water hydrating
a myoglobin surface at low hydration level ($h=0.35$~g H$_2$O/g of
protein), corresponding to a number of water molecules sufficient to
cover the entire protein surface \cite{Settles-Doster1996}.

Franzese and de los Santos in 2009 \cite{FDLS2009} found that a
water monolayer partially hydrating a hydrophobic surface, described
by the model considered here, would
display a very slow dynamics for the HBs at low $P$ and $T$. In these
conditions the HB correlation function $C(t)$, that
quantifies how much the HBs are correlated in time, is almost not changing
in time, showing that the water dynamics is completely frozen. Hence, 
water is in its glassy state at low $T$ and low $P$, consistent with
the observed freezing of the incoherent intermediate scattering function of water 
hydrating myoglobin at hydration level $h=0.34$~g H$_2$O/g of
protein at about $180$~K~\cite{Doster2010}. Glassy water is
observed also in bulk, but  below $150$~K. 

This slowing down of the dynamics is
well understood in our model where, at low $T$ and low $P$, the number of
HBs largely
increases when the $T$ decreases. This progressive building
up of the HB network traps the water molecules in 
a percolating network of HBs that leaves small areas of the
hydrophobic surface completely dehydrated.

At higher $P$ the decreased number of HB allows to 
the small dry cavities to slowly equilibrate, due to 
large-scale rearrangements of the HBs and leading to partial
dehydration of the surface. In this case water can slowly flow on the
surface because the high pressure reduces the volume per molecules and 
disfavors the formation of HBs. Hence, the HB network builds up at
lower $T$ with respect to the low pressure condition \cite{FDLS2009}. In
this case the 
long time behavior of $C(t)$ is well  described by a stretched exponential function 
\begin{equation}
C(t)=C_0 \exp\left[-\left(t/\tau\right)^\beta \right]
\label{stretched}
\end{equation}
where $C_0$, $\tau$ and $\beta\leq 1$ are fitting constant.
For $\beta=1$ the function is exponential, and the more stretched is
the function, the smaller is the exponent $0<\beta\leq 1$.

Franzese and de los Santos~\cite{FDLS2009} predicted that at low $P$,
by increasing $P$, (i) the
time needed for the HBs to decorrelate decreases, i.e. water can relax
more rapidly, and (ii)  the $\beta$
exponent decreases  going from $\beta=0.8$ to
$\beta=0.4$ for a pressure approaching a characteristic value $P_{C'}$.
We will discuss further about $P_{C'}$ in the following. 
By increasing the pressure even further, for $P>P_{C'}$, the HB
correlation function relaxes faster and the exponent becomes $\beta=1$, i.e.  $C(t)$
becomes an exponential function. 

The prediction about $\beta$ is consistent with the 
experimental findings of Settles and Doster showing that, for water hydrating
myoglobin at low hydration level, the incoherent intermediate
scattering function at large $Q$ vector,  i.e. the dynamics of
density fluctuations at short distance, are well described by 
stretched exponential functions with $\beta$ varying between $0.4$ and
$0.3$ at $320$~K \cite{Settles-Doster1996}.

We observe that the theoretical lower limit for $\beta$ is
expected to be $1/3$ \cite{Campbell88,Fierro1999} and that 
 $1-\beta$ is a measure of the heterogeneity in the system.
Therefore, the prediction~\cite{FDLS2009} that
the stretching parameter $\beta$ approaches its smallest possible value
when the pressure tends to $P_{C'}$  implies that the water monolayer
reaches its maximum in heterogeneity at $P_{C'}$.
In the following we will discuss further this point.

\subsection{Thermodynamics at very low temperature}

To understand the origin of the heterogeneity at $P_{C'}$, we recall here that 
the properties of water are consistent with theories
that propose different mechanisms and different phase behaviors at
very low temperature (approximately $150$~K$\leq T\leq 200$~K). 
At these temperatures, supercooled bulk water forms ice, while
confined water in appropriate conditions can be kept liquid \cite{Stanley2010}.
The different theories can be summarized in four possible scenarios for the
$P-T$ phase diagram.

(i) In the {\it stability limit} (SL) scenario~\cite{Speedy82} it is
hypothesized that the limits of stability of superheated-stretched
liquid water changes its slope in the $P$--$T$ plane from positive at
high $T$, to negative at low $T$ and negative $P$, giving as a
consequence the anomalous increase at low $T$ of quantities such as $K_T$, $C_P$ and
$\alpha_P$. 

(ii) In the \emph{liquid--liquid critical point} (LLCP) scenario
\cite{llcp} it is hypothesized the existence of a 
first--order phase transition line in the supercooled liquid region, 
with negative slope in the $P$--$T$
plane and  terminating in a critical point $C'$.  This phase
transition separates two liquid phases, both metastable with respect
to the crystal phases: one with low density, resembling the LDA
disordered ice,
and one with high density, resembling the HDA disordered ice. 
The low-density-liquid (LDL), high-density-liquid (HDL) critical point
$C'$ has been predicted at positive $P$ \cite{llcp}  or negative
$P$~\cite{Tanaka96}, depending on the water model adopted in
simulations.  The anomalies of water are the consequence of
approaching $C'$.

(iii) In the \emph{singularity--free} (SF) scenario~\cite{Sastry1996} it is
hypothesized that the HBs have no cooperativity. In this case it is
shown that the anomalous increase of $K_T$, $C_P$ and
$\alpha_P$ is a consequence of the 
low-$T$ anticorrelation between volume and
entropy, also responsible for negative slope in 
the $P$--$T$ plane of the line of temperatures of maximum density
(TMD).

(iv) In the~\emph{critical--point free} (CPF) scenario
\cite{Angell2008} it is
hypothesized that the LDL-HDL first--order phase transition line 
extends to $P<0$, reaching the  superheated
limit of stability of liquid water and with no critical
point. As a consequence the HDL has a superheated-stretched limit of
stability similar to that predicted in scenario (i).

\begin{figure}
(a) \hspace{10cm} (b) 
\includegraphics[width=8cm]{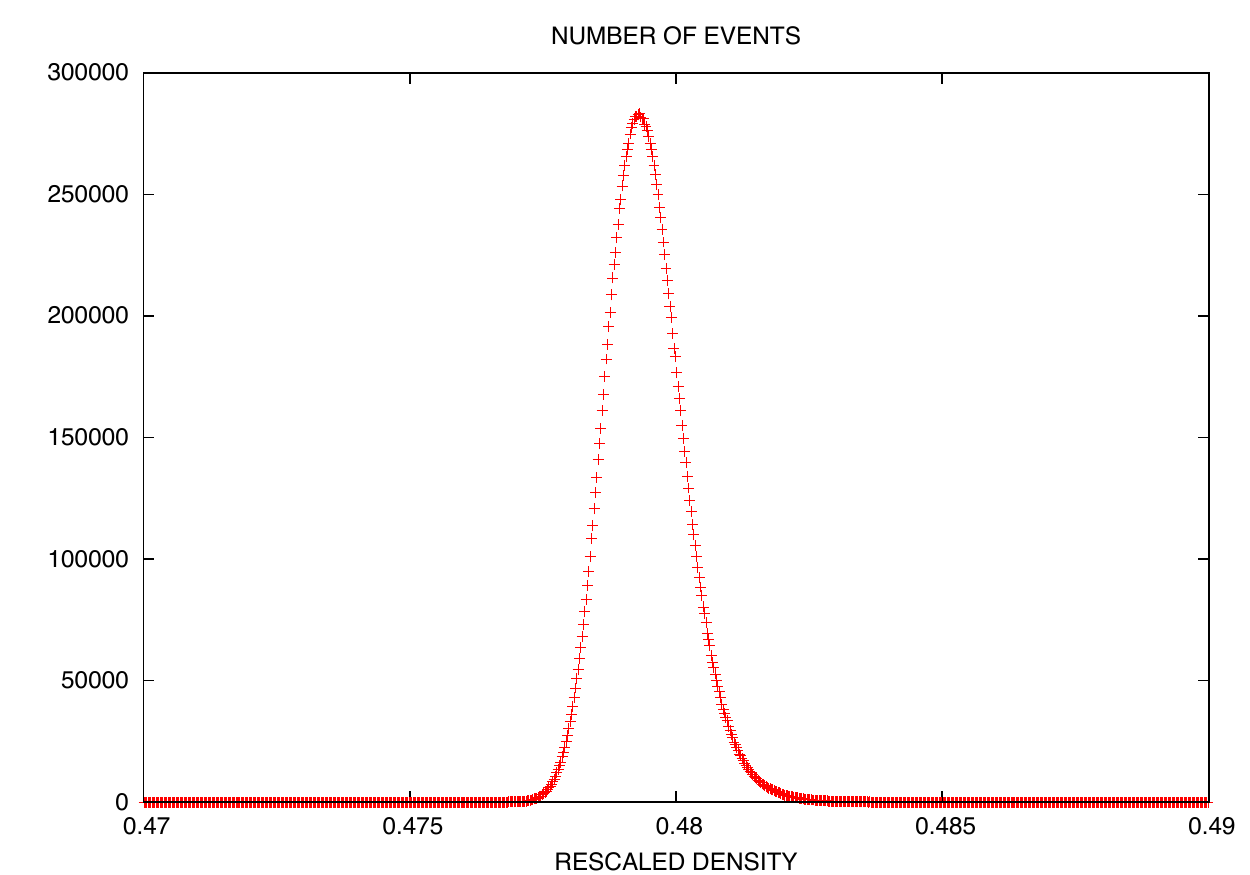}
\includegraphics[width=8cm]{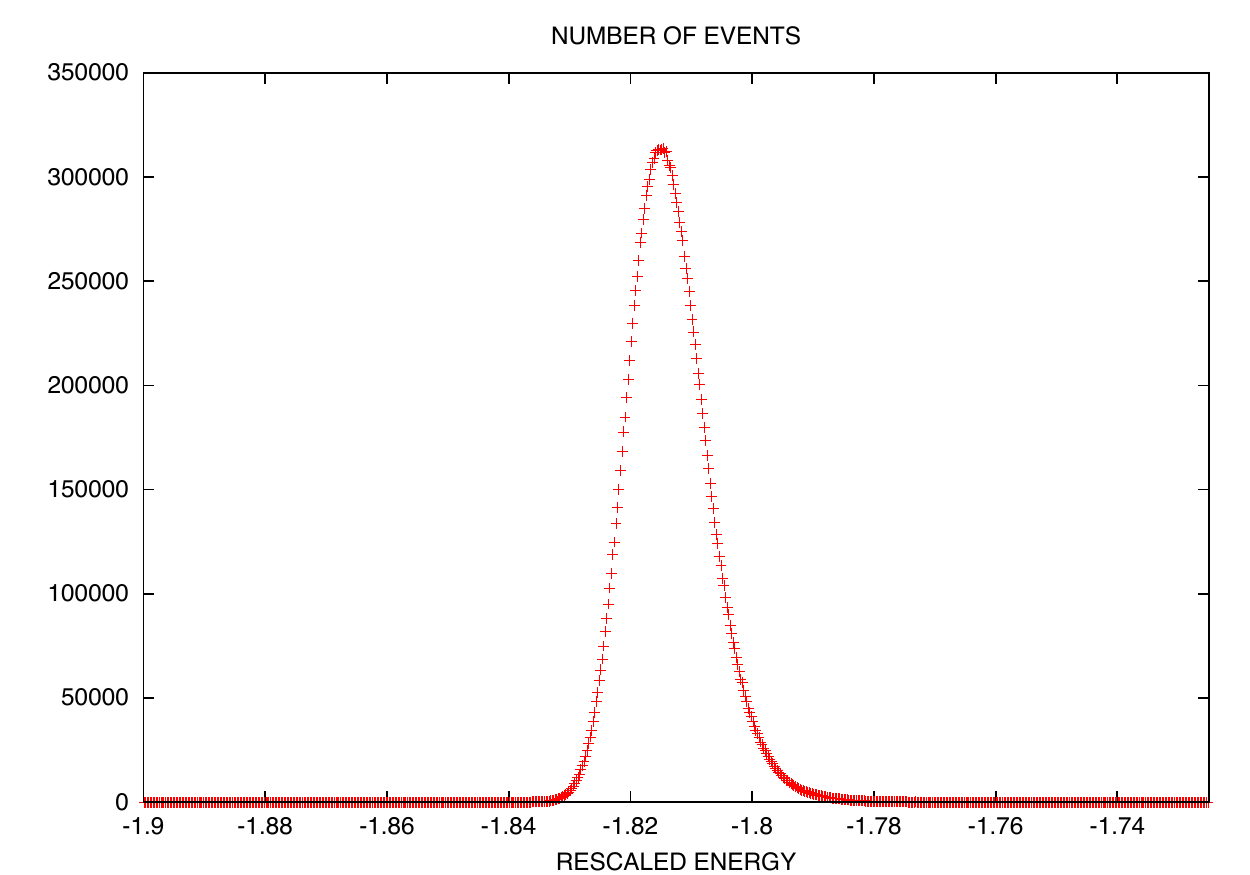}
(c) \hspace{10cm} (d)
\includegraphics[width=8cm]{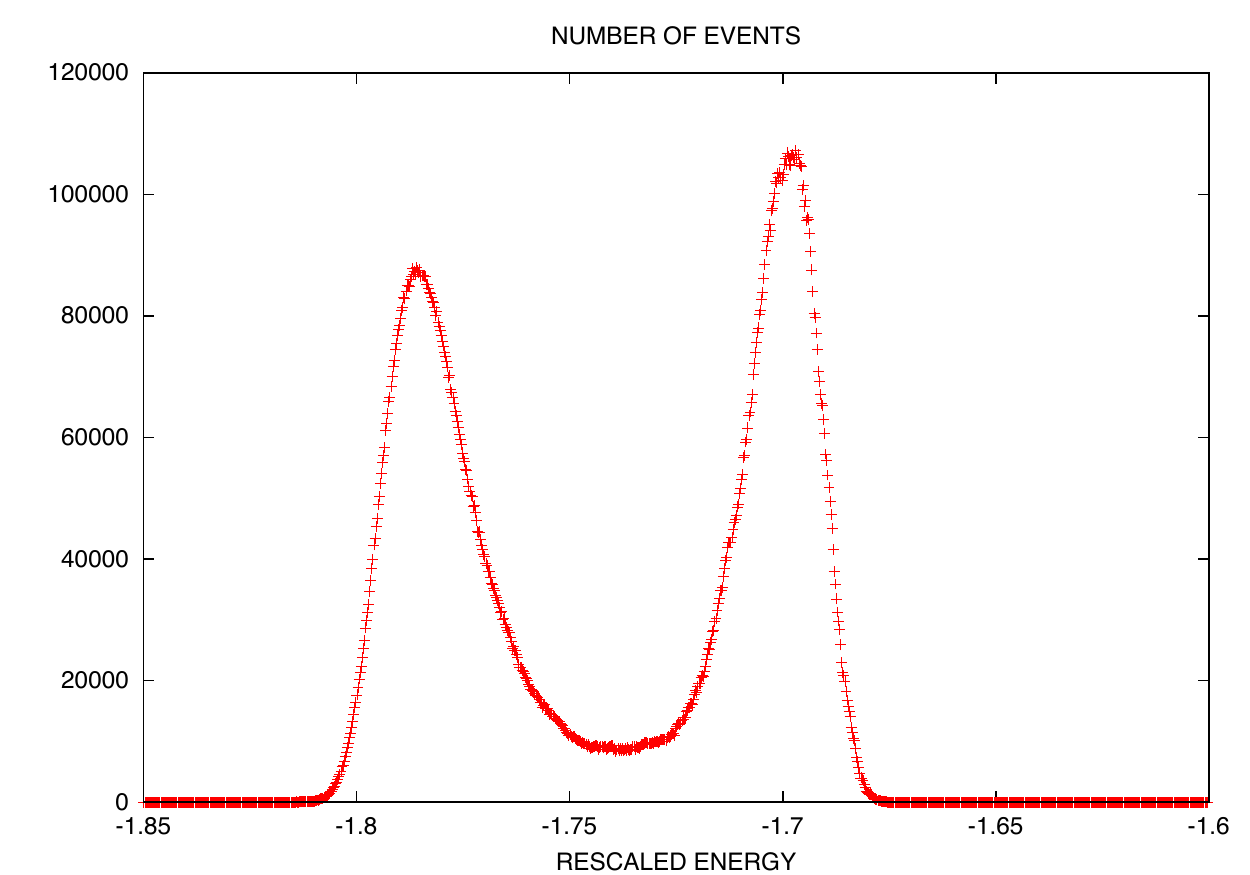}
\includegraphics[width=8cm]{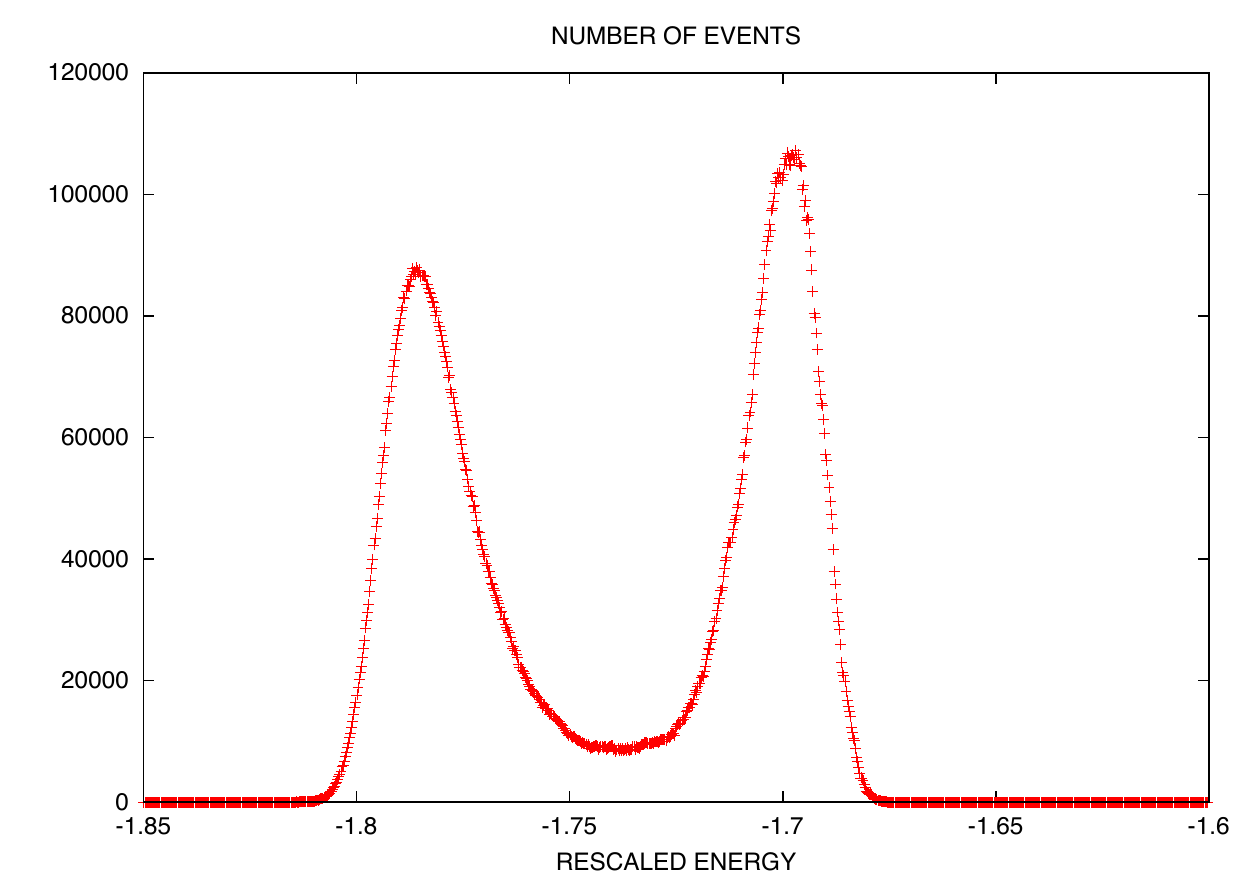}
\caption{The histogram of rescaled density $\rho v_0$ and rescaled
  energy $[U+ \mathscr{H}_{\rm HB} + \mathscr{H}_{\rm
    coop}]/\epsilon$
above [panels (a)-(b)] and below [panels (c)-(d)] the liquid-liquid critical point $C'$.
In the one-phase region, density histogram (a) and energy
histogram (b) display one single maximum.
At the liquid-liquid phase separation, density histogram (c) and energy
histogram (d) display two maxima separated by
  a minimum, corresponding to two  coexisting phases with different
  densities and energies.
Calculations are from Monte Carlo simulations of a monolayer with
$N=15625$ water molecules at
$T=175.1$~K and $P=0.12$~GPa (a)-(b), and  
$T=173.8$~K and $P=0.13$~GPa (c)-(d). 
The liquid-liquid critical point $C'$ is estimated at 
$T_{C'}\simeq 180$~K and $P_{C'}\simeq 0.13$~GPa.}
\label{histograms}
\end{figure}

As discussed by Stokely et al.~\cite{Stokely2010},  it is
still unclear which of the scenarios best describes water, because
there is no definitive experimental test.   In
Ref.~\cite{Stokely2010},  the same tractable model 
presented in the previous section
is analyzed by means of theoretical calculations and numerical
simulations, showing that the four scenarios (i)--(iv) 
may be mapped in the space of the parameters $J$ and $J_\sigma$,
representing the strength of the HB directional component and 
the strength of the HB cooperative component, respectively.
The relation $J_\sigma/J=1/10$ discussed at the end of the previous
section, and based on estimates from experimental data, supports
the prediction of a liquid--liquid critical point $C'$ at positive pressure
for supercooled water (Fig.\ref{rho}b). The model allows to
distinguish the LDL phase and the HDL phase by their different densities
and characteristic energies. The presence of two maxima in the
distributions of these quantities marks the coexistence of the two
phases and corresponds to the occurrence of two minima in the free
energy of the system (Fig.\ref{histograms}). 

At the critical point $C'$ the cooperativity of the HBs is
maximum. Hence, cooperative rearrangements of the HB are necessary 
to allow the relaxation of the dynamics. These rearrangements occur on
different length-scales, each associated with a different time-scale.
As a consequence a single time-scale cannot
be defined for the dynamics, resulting in a stretched decay of the correlation
function, i.e. in heterogeneous dynamics.
It is, therefore, $C'$ the origin of the heterogeneity described in
the previous section and occurring at $P_{C'}$ of the liquid--liquid critical
point.

\subsection{Hydration percolation}

It possible to study the cooperative regions, and their length-scales,
by using a geometrical approach based on the concept of {\it correlated
percolation} \cite{Franzese1998,MSSSF2009}. We define
a cluster of correlated water molecules as described in the following
steps. 
\begin{itemize}
\item
The first step consists in including in the cluster one of the
bonding indices $\sigma_{i,j}$ of a randomly
selected water molecule $i$ in the hydration shell. 

\item
The second step is to
add to the cluster another bonding index of the same molecule $i$ with probability $p_{\rm same}\equiv
\min\left\{1,1-\exp[-J_\sigma/(k_BT)]\right\}$ 
(where $k_B$ is the Boltzmann constant), 
or the facing bonding index $\sigma_{j,i}$
of the nearest neighbor molecule $j$  with probability 
$p_{\rm facing}\equiv \min\left\{1,1-\exp[-J'/(k_BT)]\right\}$
where $J'\equiv J-P v_{HB}$. In this expression, $J'$ 
 is the $P$--dependent effective coupling 
between two facing indices as results from the enthalpy 
$U+ \mathscr{H}_{\rm HB} + \mathscr{H}_{\rm coop}  + PV$ of the
system Eq.s~(\ref{U0})-(\ref{coop}).
The quantity $J'$ can be positive or negative depending on $P$.
If $J'>0$, an index can be added, with probability $p_{\rm facing}$,
to the cluster only if it is in the same state as the other indices
already in the cluster. Instead, if $J'<0$, the index can be added
only if it is in a different state with respect to the index to whom it will
be connected.

\item
The third step is to randomly select a bonding indices on the border of the
cluster and pick at random
one of the indices on the same molecule, or the facing index on a bonded
molecule, that is not already in the cluster, and to include it in the
cluster with probability $p_{\rm same}$ or $p_{\rm facing}$,
respectively. This step is repeated until all the possible directions
of growth for the cluster have been considered.
\end{itemize}

\begin{figure}
\begin{center}
\includegraphics[width=15cm]{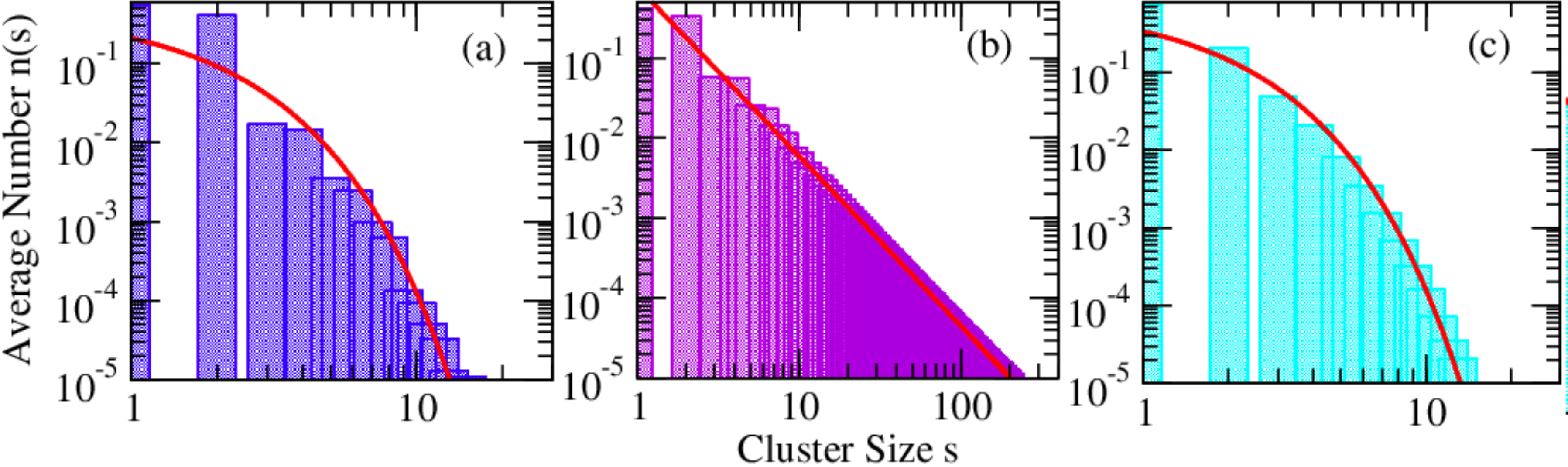}
\end{center}
\caption{Histograms of number $n(s)$ of  clusters of size $s$ near the
  liquid-liquid critical point $C'$ for a monolayer with $N=25600$
  water molecules. For $T<T_{C'}$ (a) and $T>T_{C'}$ (c) the distribution of $n(s)$ is
  exponential, while for $T\simeq T_{C'}$ (b) it becomes approximately a
  power law, whose leading term is $n(s)\simeq s^\tau$ with an
  estimate $\tau\simeq 2$. Lines are fits of the histogram with
  exponential functions in (a) and (c) and with $n(s)\simeq s^\tau$ in
(b).}
\label{ns}
\end{figure}

The resulting cluster statistically represents the region of
correlated HBs, as can be shown \cite{Cataudella1996}, and its
characteristic linear size statistically corresponds to the
correlation length of the water molecules. Therefore, in the vicinity
of the liquid-liquid critical point, where the correlation length
increases, it is possible to observe that the size of the clusters of correlated water molecules 
increases. At the liquid-liquid critical point the
correlation length diverges and a cluster of correlated water molecules 
spans (percolate) the whole monolayer. General results on correlated
percolation theory allow to find mathematical relations between 
thermodynamics quantities and percolation quantities. In particular, 
it can been shown that the mean size of the clusters defined above
diverges with the same power--law exponent as 
the compressibility of water and that the distribution of number $n(s)$ of finite cluster of
size $s$ per water molecule is exponential far from the
liquid-liquid critical point $C'$, while follows a power law with
exponent $\tau$ near $C'$. From general considerations it is possible
to show that $\tau=1+d/D_F$ where $d=2$ is the effective
dimensionality of the monolayer and 
$D_F$ is the fractal dimension of the clusters \cite{Franzese1996}. Preliminary estimate of
$\tau\simeq 2$ suggests that the clusters of  correlated water
molecules are compact with $D_F\simeq 2$ (Fig.\ref{ns}) \cite{bianco2010}. 
Since the
compressibility is proportional to the density fluctuations, the
clusters allows for a geometrical analysis of the diverging density
fluctuations near the liquid-liquid critical point \cite{Franzese2010}.

\subsection{Hydrogen bonds dynamics on Hydrated Protein}

The density fluctuations are observable also far from the
liquid-liquid critical point $C'$. In particular,  they can be
observed along a line in the $P$--$T$ 
phase diagram that emanates from $C'$
into the one-phase region and marks the maxima of the correlation
length. This line has been named after Widom 
\cite{Xu2005,FS2007,Kumar05062007} and can be characterized in the
study of hydrated protein. For example, 
Kumar et al. \cite{KFS2008} used the tractable model 
described above to investigate the case of 
a percolating monolayer of water molecules adsorbed on a protein
surface, with hydration level about $h\simeq 0.4$ g H$_2$O/g dry
protein. Under these conditions the protein is immobile and  inhibits
the ice crystallization because it forces the water molecules out of
the positions corresponding to a crystal configurations.
The authors studied the HB dynamics, regardless if the HBs are formed
within the water molecules or with the surface.

They first locate the Widom line, by  Monte Carlo simulations, and
observed that it corresponds to the locus where there is the largest
change in the number of HBs. At $P$
and $T$ above the Widom line, water has a few HBs, while at 
$P$ and $T$ below the Widom line, it has a well-developed network of
HBs. This change of structure is reflected by a maxima in constant-$P$
the specific heat along the Widom line. The simulations also reveal
the presence of 
a dynamic crossover in the HB correlation function $C(t)$ when the
Widom line is crossed  at constant $P$. Using mean field theory and
making the hypothesis that the dynamics is dominated by the
rearrangements of the HBs,
the authors calculate the activation energy for the relaxation of the
system and show that it gives the same relaxation time calculated by
Monte Carlo simulations  \cite{KumarFS2008}.
The proposed mechanism  consists in breaking a HB that does not fit
into the tetrahedral arrangement and reorient the molecule to optimize
locally  the tetrahedral  configuration. This mechanism 
has been confirmed also by simulations of the hydration shells of
elastin-like and collagen-like peptides \cite{Vogel2009}.
In particular, Kumar et al. \cite{KumarFS2008}  predict
(i) how this barrier is affected by the variation of $P$, 
(ii) how the  crossover $T$ is affected by the variation of $P$, 
and (iii) that
for any $P$ the correlation time at the crossover $T$ is the same
(isochronic crossover). Experiments with hydrated lysozyme, spanning a
range of pressures going from ambient pressure up to 1600 bar, performed by the group of
S.-H. Chen at MIT, have confirmed these three predictions
\cite{Chu:2009hl,FranzeseSCKMCS2008}.

More recent analysis of lysozyme proteins at a lower hydration level
($h=$0.3 g H$_2$O/g dry protein) reveals another surprising results
\cite{Mazza:2009}.
At this very low hydration, dielectric spectroscopy, probing the proton
relaxation, displays that at ambient $P$ not only there is at
about 250~K the dynamic crossover described above, but also
another crossover at about 180~K. The study of the tractable model
presented here associates this lower-$T$ crossover to the saturation
of the  
cooperative ordering of the HB network. Specifically, the HBs
rearrange to maximize the number of tetrahedral orientations among the
bonds. This ordering is marked by a new specific heat maxima that can
be calculated in the model at about  180~K. Therefore, summarizing,
the model predicts a broad specific heat maximum at about 250~K, due
to the saturation 
of a macroscopic HB network, and a sharper specific heat
maximum at about 180~K, due to the saturation of the tetrahedral ordering of the HBs. By
increasing the pressure of the hydrated protein, Mazza et al. predict
that these two maxima merge and then diverge at the liquid-liquid
critical point $C'$ \cite{Mazza:2008}.

\section{Discussion: Implications in food
  science} 

The presence of a liquid-liquid critical point $C'$ at low $T$ in confined
water could be an undesirable property for the storage of frozen food and, more in
general, biological cells. This is because in the vicinity of a
critical point between two liquid phases, both metastable with respect
to the crystal phase, large density fluctuations occur. These
enhanced fluctuations would drastically
changes the pathway for the formation of a crystal nucleus, because
the crystal would form from the dense fluid, instead that from the
low density fluid. As a consequence, there would be  a strong
reduction of the crystal nucleation  
free-energy barrier and, hence, an increase by many orders of magnitude
of the crystallization rate, as theoretically  predicted by  ten Wolde
and Frenkel \cite{tenWolde1997}.  

Under this conditions the enhanced formation of ice
could destroy the cells as a consequence of the increase of volume of ice
with respect to the liquid water \cite{Sun2003,Petzold2009}.
Hence, the best way to preserve the cells would be
to freeze them at a $T$ that is far away from the liquid-liquid
critical temperature. Nevertheless, further analysis show that the
situation is even more complex. Indeed, simulations of a model with a
metastable liquid-liquid critical point display enhanced
crystallization rate not only in the vicinity of the critical point,
but in the vicinity of the whole region of liquid-liquid coexistence
\cite{Franzese2002} and possibly also in the one-phase region above
the critical point along the Widom line \cite{FS2007}. Work is in
progress to elucidate these implications.

In the evaluation of these effects is extremely important 
to properly include the interaction with the confining
surfaces. We are presently studying how to incorporate these effects
in our tractable model. A first step in this direction is to
include a description of the hydrophobic effect.
Frank and Evans \cite{frank1945} and Silverstein et al. \cite{silverstein1999}
proposed that supercooled water forms highly structured ``ice-like''
regions in the hydration shell of nonpolar solutes. 
Stillinger \cite{Stillinger1973} proposed that HBs in the
hydration shell are not significantly perturbed near small hydrophobic
solutes, while the HB network is strongly affected by hydrophobic particles
with size above a characteristic value. Chandler estimated this value of the
order of 1~nm on the basis of free energy calculations
\cite{Chandler2005}.  Muller explained the vibrational and NMR
spectroscopy 
results by suggesting enthalpic strengthening of the hydration HBs
with a simultaneous entropy increase in the hydration shell \cite{Muller1990}.
We are presently including the  enthalpic strengthening  in our
model, and properly accounting for the entropy increase for the study
of water in confined by hydrophobic nanoparticles. Our preliminary 
results \cite{Elena2010} show a surprising change of 
thermodynamic fluctuations at low $T$, whose implication is a large
decrease of compressibility also at very low nanoparticles
concentration. This finding suggests that  adding hydrophobic
particles at low concentration in organic solutions would decrease the
density fluctuations and the formation of ice.

\begin{figure}
(a) \hspace{2.8cm}
(b) \hspace{2.8cm}
(c) \hspace{2.8cm}
(d) \hspace{2.8cm}
(e) \hspace{2.8cm}
\includegraphics[width=3cm]{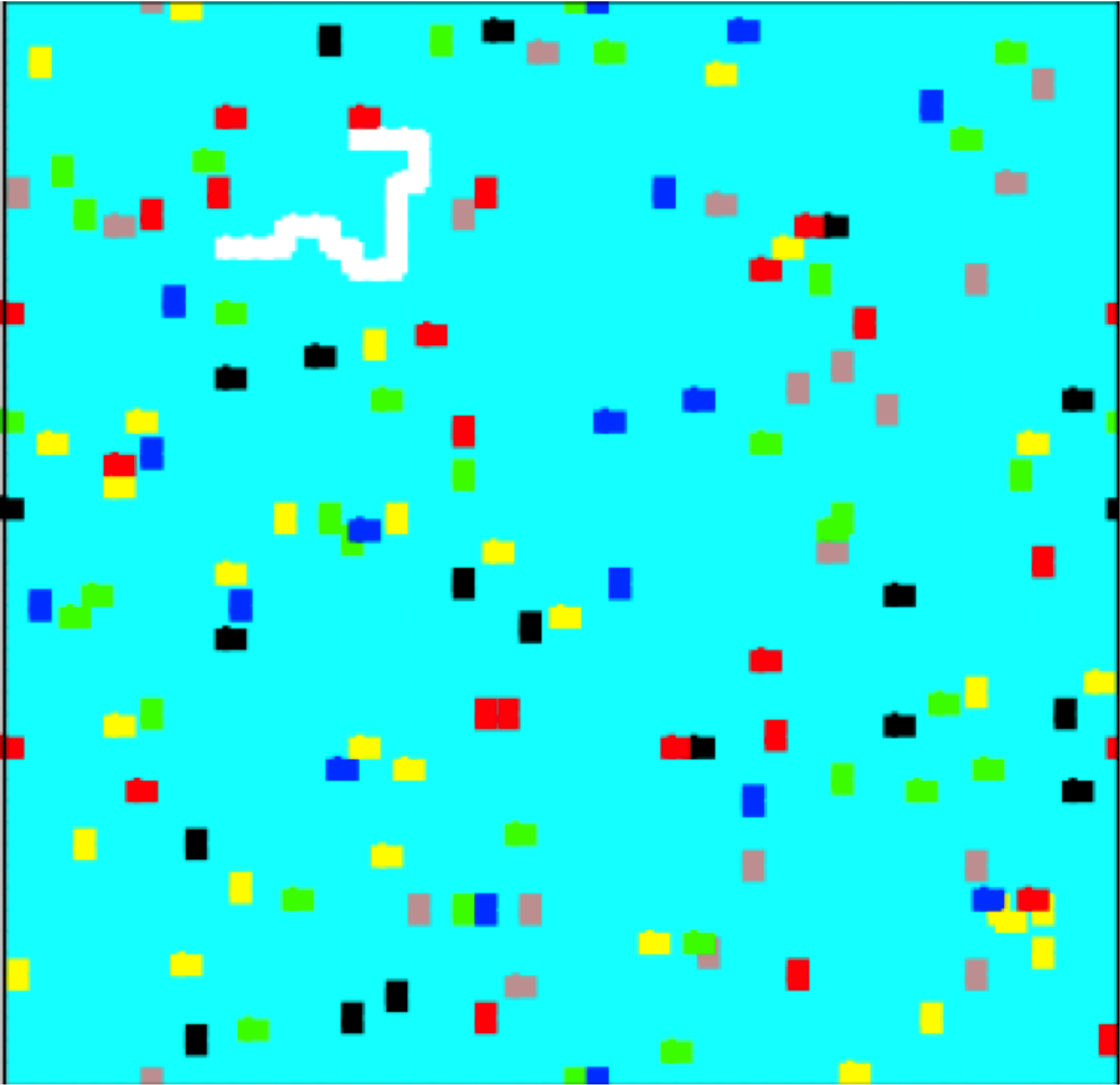}
\includegraphics[width=3cm]{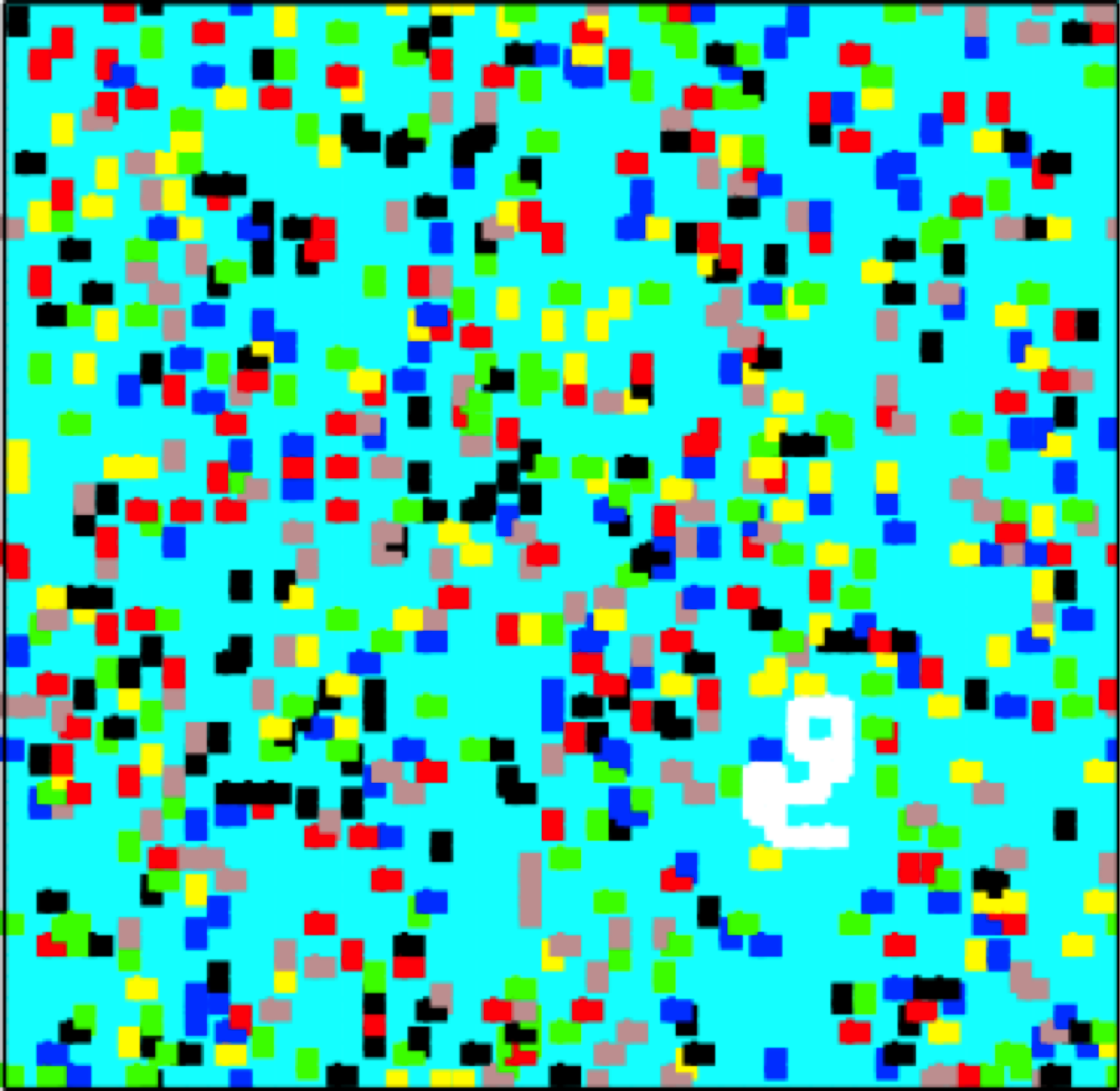}
\includegraphics[width=3cm]{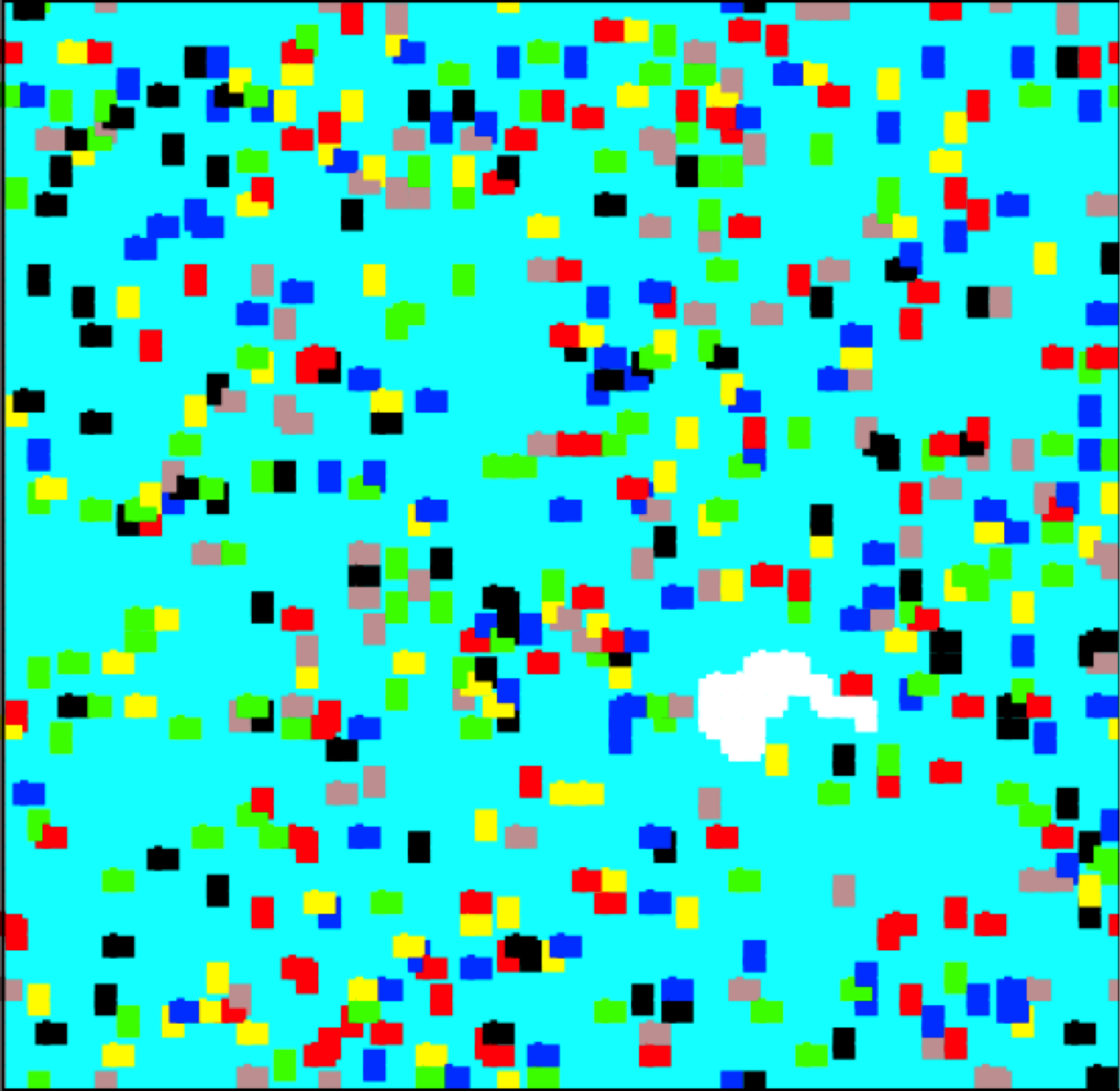}
\includegraphics[width=3cm]{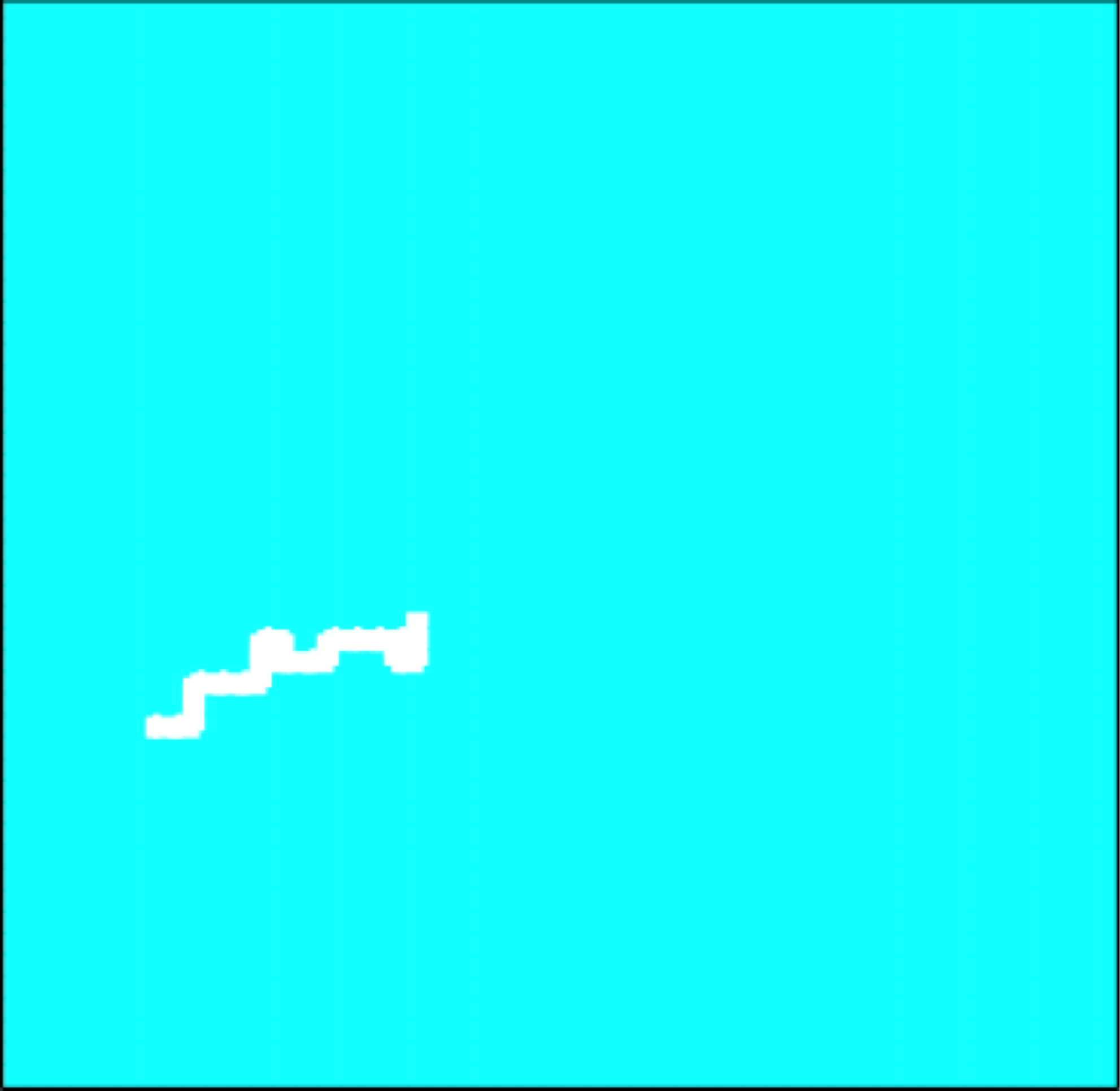}
\includegraphics[width=3cm]{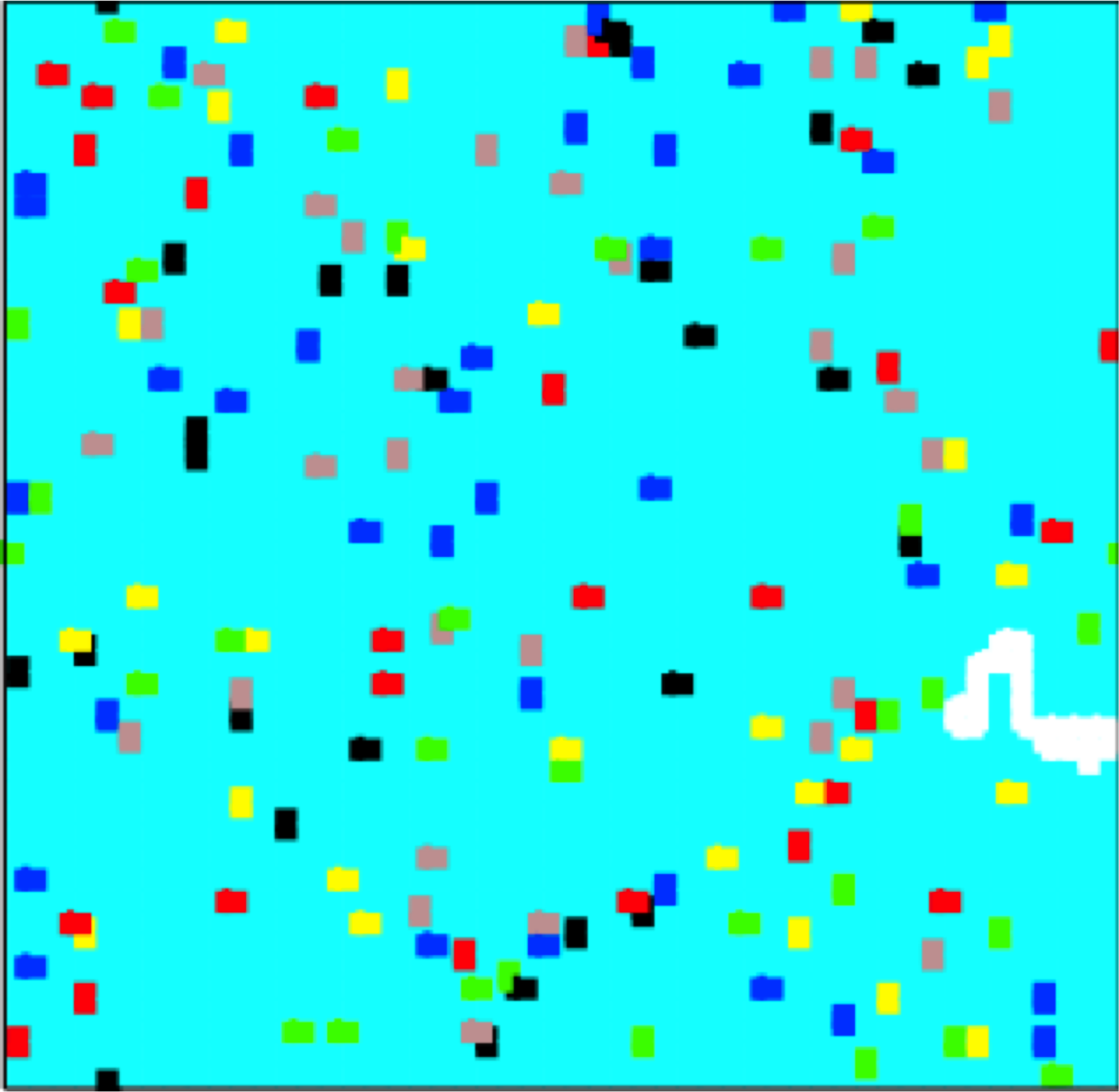}
\caption{Folding and unfolding 
of a coarse-grained protein suspended
  in water at different 
temperatures $T$ and high pressures $P$. Typical
  configurations in a system in two dimensions are represented for the
  sake of schematic description of the process.
The protein is
  represented as a fully hydrophobic chain (in white), surrounded by
  water molecules (turquoise background).  (a) At high pressure and
  high $T$, the protein is unfolded and the surrounding water has only
  a few hydrogen bonds (HBs) formed, represented as colored sticks. 
Different colors of the HBs correspond to different relative
orientations of the HBs. (b) At the same pressure but lower $T$, the
protein start to fold in a {\it molten globule} state. (c) At lower
$T$ the protein folds, while the surrounded water has a large number
of HBs. (d) At much lower $T$ we observe {\it cold denaturation} of
the protein when the number of water HBs is largely reduced due to the
combined action of $T$ and $P$. (e) At higher $P$ the denaturation effect is
observed at higher $T$, and we observe the protein only in open configurations.
%$P=0.8$~GPa, $T=141$~K, $T=1878$~K, $T=3622$~K, 
%$P=1.0$~GPa, $T=832$~K,
%$P=2.0$~GPa, $T=1878$~K
}
\label{protein}
\end{figure}

Further confirmation of the validity of our assumption about the
hydrophobic effect comes from 
another study that we are performing to establish if our tractable
model is able to describe the stability of proteins with respect to
changes of temperature and pressure. In this study we consider how the
hydrophobic interaction with water of a coarse-grained protein induces
hot denaturation, folding, cold denaturation and pressure
denaturation. Our preliminary results \cite{Svilen2010} display a
region of stable folded configurations that, in the $P$--$T$ phase
diagram, has the same qualitative features of the experimental
stability diagram of  myoglobin \cite{Smeller2002} (Fig.\ref{protein}).

\section{Conclusions}

We have introduced a tractable model for a water monolayer hydrating
surfaces of proteins and, more generally, confined water. The model
includes HB cooperativity and elucidates how the many-body component of
the HB is important to understand the low-$T$ behavior of water. In
particular, parameters estimated from the experiments suggest the
occurrence of a critical point at $T\simeq 180$~K and $P\simeq
0.13$~GPa at the end of a first order coexistence  line between two
liquids with different densities \cite{FS2002,FMS2003,Stokely2010}.

The model shows that 
the liquid-liquid critical point affects the low-$T$ dehydration of a hydrophobic
surface  \cite{FDLS2009}. The cooperativity of water induces dynamic
heterogeneities that reach their maximum when the cluster of
correlated HBs percolate \cite{MSSSF2009,Franzese2010,bianco2010}.

This heterogeneous dynamic behavior is revealed by a strongly
non-exponential relaxation of the HB dynamics and by a 
subdiffusive translational motion of the water
molecules in the hydration shell \cite{FDLS2009}, as observed in hydrated proteins at low 
$T$ \cite{Settles-Doster1996,Doster2010}.

The model predicts that, at low protein hydration,
shell water should be
characterized by two structural transitions. One associated to the
macroscopic formation of HBs \cite{KumarFS2008,KumarFBS2008,KFS2008},
occurring at about 250~K for low-hydrated lysozyme, and another
associated to the tetrahedral reordering of the HBs, at about 180~K
\cite{Mazza:2009}.
These two structural changes are at the origin
of two dynamic crossovers: the one at higher $T$ has been observed by
QENS experiments on hydrated lysozyme
\cite{Chu:2009hl,FranzeseSCKMCS2008}, and both have been measured by
dielectric spectroscopy  on hydrated lysozyme \cite{Mazza:2009}.
The pressure behavior of these crossover is consistent with the
presence of the liquid-liquid critical point $C'$ at high $P$ and low
$T$ \cite{FranzeseSCKMCS2008,Mazza:2008}. 
 
The large increase of  density fluctuations in the vicinity  of the
critical point $C'$ is expected to enhance the water crystallization
process. As a consequence of the expansion of ice, this process could
destroy biological structures in a crowed environment, as for example
in food stored at low temperature \cite{Sun2003,Petzold2009}. It is, therefore, relevant to
understand if this process is affected by confinement or if it 
could be controlled. Our preliminary
results \cite{Elena2010} show that hydrophobic confinement has a
strong effect of the thermodynamics of water, suppressing the density
fluctuations associated to the critical point $C'$. This result
suggests that by dissolving hydrophobic nanoparticles  at low
concentration it could be possible to control the water compressibility
and the formation of large crystals.

These predictions are based on a modelization  of the hydrophobic
interaction that is able to reproduce a stability diagram for a
coarse-grained protein. In particular, the model shows that the
protein cold denaturation and the pressure denaturation 
can be explained as a consequence of the strengthening of interfacial water-water
HBs at the hydrophobic interface and the proper account of the entropy
change due to the presence of the interface \cite{Svilen2010}.

Work is in progress to include other features in the model and to use it
to make other predictions in different contexts. For example, we are
extending the model in such a way to describe the ice formation and
analyze how the crystallization process is affected by
interfaces. We are also developing the generalization in three
dimensions to allow the study of many layers of hydration water and to
extend the investigation to the bulk case. In this way our research about
protein stability will be developed also in bulk water. All together
these generalizations will allow us to explore situations of possible
interest in food science and biology, as for example the effect on
water structure and dynamics of preservative agents such as trehalose,
or antifreeze proteins, or cryoprotectants such as glycerol or
dimethyl sulphoxide. 

\section*{Acknowledgments}

G.F. thanks for collaboration and helpful discussions
M. C. Barbosa,
S. V. Buldyrev,
F. Bruni,
S.-H. Chen,
A. Hernando-Mart\'{\i}nez,
P. Kumar, 
G. Malescio,
F. Mallamace,
M. I. Marqu\'es,
M. G. Mazza,
A. B. de Oliveira,
S. Pagnotta,
F. de los Santos,
H. E. Stanley, 
K. Stokely,
E. G. Strekalova,
P. Vilaseca,
and the Spanish Ministerio de Ciencia e Innovaci\'on
Grants FIS2009-10210 (co-financed FEDER)
for support.

\end{document}